\documentclass{ws-p9-75x6-50}

\begin{document}

\title{Spectral Structure of Quantum Line
with a Defect}

\author{
Taksu Cheon$^1$
Tam{\'a}s F{\"u}l{\"o}p$^2$
Izumi Tsutsui$^3$
}

\address{
${}^1$Kochi University of Technology,
Tosa Yamada, Kochi 782-8502, Japan
\\E-mail: taksu.cheon@kochi-tech.ac.jp
\\
${}^2$Roland E\"{o}tv\"{o}s University,
P\'{a}zm\'{a}ny P. s\'{e}t\'{a}ny 1/A, 
H-1117 Budapest, Hungary
\\E-mail: fulopt@poe.elte.hu
\\
${}^3$High Energy Accelerator Research Organization (KEK),
Tsukuba 305-0801, Japan\\E-mail: izumi.tsutsui@kek.jp
}  

\maketitle

\abstracts{
We study the spectral properties of one-dimensional
quantum wire with a single defect.
We reveal the existence of the non-trivial topological 
structures in the spectral space of the system, which are
behind the exotic quantum phenomena that have
lately been found in the system.
}
With the progress of the nanotechnology,
it has become possible 
to manufacture quantum systems with desired specification.\cite{TH99}
The theoretical study of simple quantum system 
with nontrivial properties
is now a legitimate
and relevant subject in wider context
outside of mathematical physics. 
It has been lately pointed out that 
one of such simple model systems
of the idealized quantum wire with a single defect \cite{SE86,AG88}
possesses the properties such as 
strong vs weak coupling duality and spiral spectral 
anholonomy,\cite{CH98,CS99}
the features usually associated with the non-Abelian gauge field 
theories.
Despite its simplicity, the model is  a very generic
one in the sense that it represents the 
long wave-length limit of arbitrary one-dimensional
potential with finite spatial support.
As such, probing those 
phenomena in its precise working is worthwhile, if only
for its mathematical feasibility. 
That is exactly what we attempt in this paper.

We consider a quantum particle in 
one-dimensional line with a single defect
placed at $x = 0$.
In formal language, the system is described by
the Hamiltonian
\begin{eqnarray}
\label{F1}
H = -{{\hbar^2}\over{2m}}{{d^2}\over{d x^2}} ,
\end{eqnarray}
defined on proper domains in the Hilbert space
${\cal H} = L^2({\bf R}\setminus\{0\})$.
We ask what
the most general condition at $x=0$ is.
%
We define the two-component vectors, \cite{FT00}
\begin{eqnarray} 
\label{F2} 
\Phi =
  \left( {\matrix{{\varphi (0_+)}\cr
                  {\varphi (0_-)}\cr}
         } 
  \right),
\qquad 
\Phi' =
  \left( {\matrix{{ \varphi' (0_+)}\cr
                  {-\varphi' (0_-)}\cr}
         } 
  \right),
\end{eqnarray}
from the values and derivatives of a wave function
$\varphi(x)$ at the left $x = 0_-$ 
and the right $x = 0_+$ of the missing point.
The requirement of self-adjointness of the Hamiltonian
operator (\ref{F1}) is satisfied if
probability current 
$j(x) = - i\hbar(
(\varphi^*)'\varphi - \varphi^* \varphi' )/(2m)$
is continuous at $x = 0$. 
In terms of $\Phi$ 
and $\Phi'$, this requirement is expressed as
\begin{eqnarray} 
\label{F2a}
\Phi'^\dagger \Phi - \Phi^\dagger \Phi' = 0 ,
\end{eqnarray}
which is equivalent to
$|\Phi-i L_0 \Phi'|$  $=$ $|\Phi+i L_0 \Phi'|$
with $L_0$ being an arbitrary constant in the
unit of length.  
This means that, with a two-by-two unitary matrix
$U\in U(2)$, we have the relation,
\begin{eqnarray} 
\label{F3} 
(U-I)\Phi+iL_0(U+I)\Phi'=0
\ .
\end{eqnarray}
This shows that the entire family $\Omega$ of
contact interactions admitted in quantum mechanics
is given by the group $U(2)$.
In mathematical term, the domain
in which the Hamiltonian $H$ becomes self-adjoint
is parametrized by $U(2)$ --- there is
a one-to-one correspondence between
a physically distinct contact
interaction and a self-adjoint Hamiltonian.
We use the notation $H_U$
for the Hamiltonian with the contact interaction specified 
by $U \in \Omega$ $\simeq U(2)$.

We now consider following {\it generalized parity} 
transformations:
\begin{eqnarray} 
\label{P1}
& &
{\cal P}_1: \, \varphi (x) 
\rightarrow  
({\cal P}_1\varphi)(x) := \varphi( - x),
\\
& &
{\cal P}_2: \, \varphi (x) 
\rightarrow  
({\cal P}_2\varphi)(x) := 
 i[\Theta(-x) - \Theta(x)]\varphi(-x)\ .
\\
& &
{\cal P}_3: \, \varphi (x) 
\rightarrow  
({\cal P}_3\varphi)(x) := [\Theta(x) -
\Theta(-x)]\varphi(x)\ .
\end{eqnarray}
These transformations satisfy the anti-commutation relation
%
\begin{eqnarray} 
\label{PR}
{\cal P}_i {\cal P}_j 
= \delta_{ij}+i\epsilon_{ijk} 
{\cal P}_k .
\end{eqnarray}
Since the effect of ${\cal P}_i$ on the boundary vectors $\Phi$ 
and $\Phi'$ are given by
$
\Phi 
\buildrel {\cal P}_i \over \longrightarrow \sigma_i \Phi\ , 
$
$
\Phi' 
\buildrel {\cal P}_i \over \longrightarrow \sigma_i \Phi'\ ,
$
where $\{ \sigma_i \}$ are the Pauli 
matrices,  
the transformation ${\cal P}_i$ on an element $H_U$$\in \Omega$
induces the unitary transformation 
\begin{eqnarray} 
\label{P3}
U 
\buildrel {\cal P}_i \over \longrightarrow \sigma_i U \sigma_i
\end{eqnarray}
on an element $U$ $\in$ $U(2)$.
The crucial fact is 
that the transformation ${\cal P}_i$ turns one 
system belonging to $\Omega$ into another 
one with {\it same spectrum}.  
In fact, with any ${\cal P}$ defined by
$
{\cal P}
:= \sum_{j = 1}^3 c_j \, {\cal P}_j
$
%
with real 
$c_j$ with constraint $\sum_{j = 1}^3 c_j^2 = 1$,
one has a transformation 
\begin{eqnarray} 
\label{P8}
{\cal P} H_U {\cal P} 
=H_{U_{\cal P}} 
\end{eqnarray}
where $U_{\cal P}$ is given by
%
\begin{eqnarray} 
\label{P9}
U_{\cal P} :=    \sigma U \sigma
\end{eqnarray}
with $\sigma:=\sum_{j = 1}^3 c_j\, \sigma_j$.
One sees, from (\ref{P8}), that the system described by
the Hamiltonians $H_U$ has a family of systems
$H_{U_{\cal P}}$ 
which share the same spectrum with $H_U$. 
%
%
\begin{figure}[t]
\ \ \
\epsfxsize=14pc \epsfbox{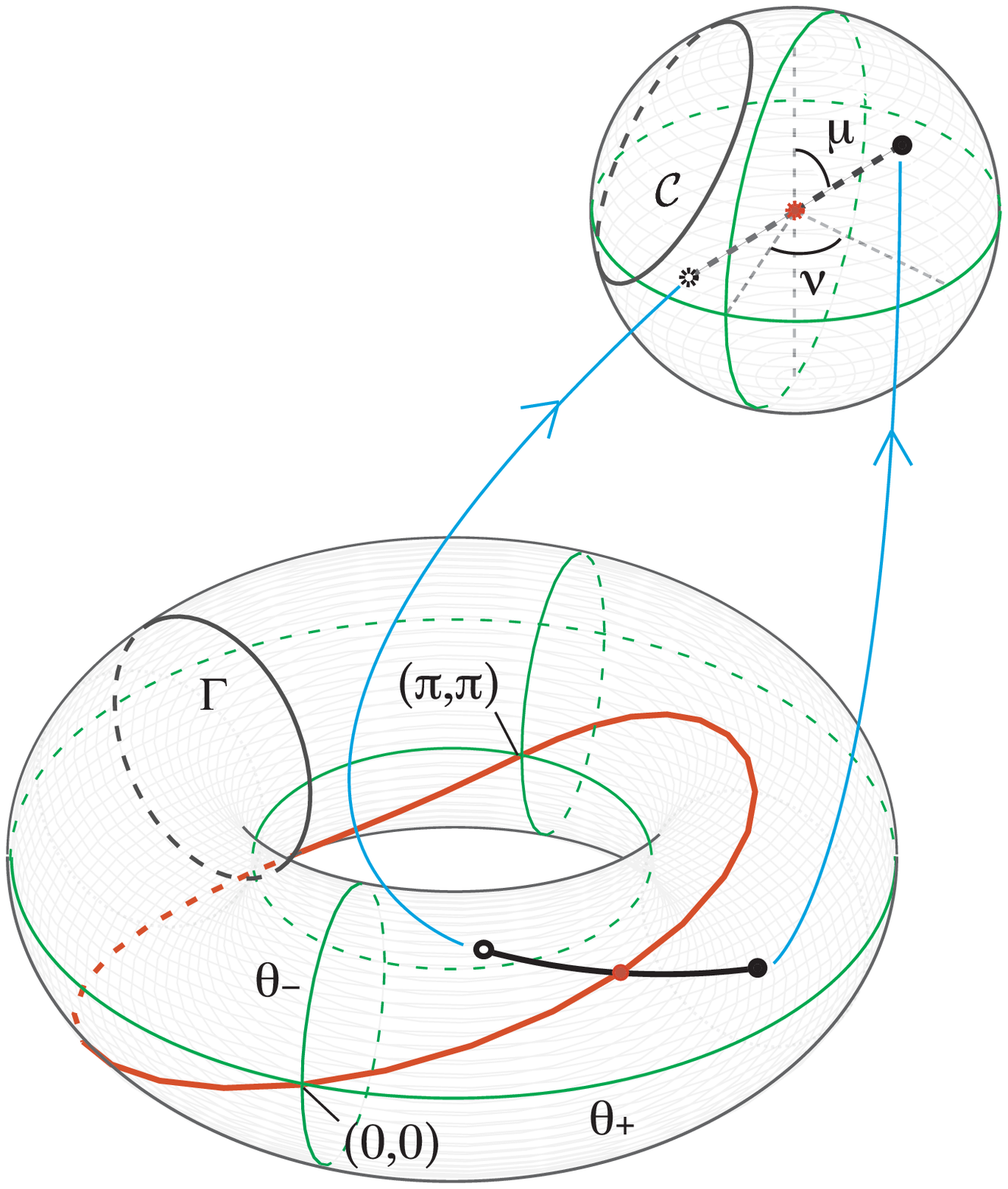}
\ \ \ \ \ \ 
\ \ \ \ \ \ 
\ \ \ \ \ \ 
\epsfxsize=11pc \epsfbox{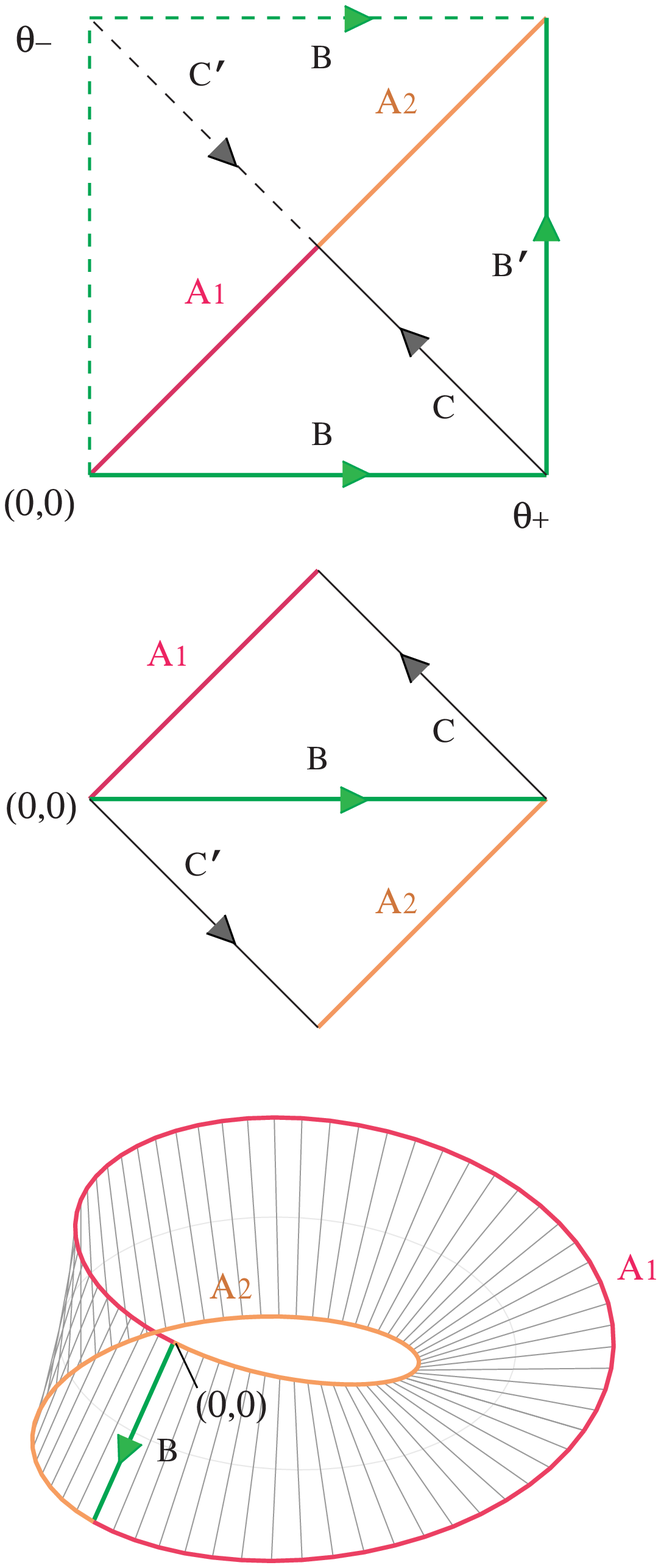}
\caption{
(a) [left] :
The parameter space
$\{( \theta_+, \theta_-, \mu, \nu)\}$ is
a product of the spectral torus $T^2$ 
specified by the angles
$(\theta_+, \theta_-)$
and the isospectral sphere $S^2$ 
specified by the angles
$(\mu, \nu)$.
(b) [right] :
In the top figure, the distict spectral space $\Sigma$
is the triangle surrounded
by edges $A_1 + A_2$, $B$ and $B'$.
Since a subtriangle
is spectrally identical to its isospectral
image $B$--$C'$--$A_2$, $\Sigma$ can be represented
by the square $A_1$--$C'$--$A_2$--$C$ in the middle figure.
When the two spectrally identical edges
$C$ and $C'$ are stitched together with the right orientation,
we obtain
the M{\"o}bius strip with boundary $A_1$--$A_2$
(the bottom figure).
}
\end{figure}

Let us suppose that the matrix $U$ is diagonalized with
appropriate $V \in SU(2)$ as
\begin{eqnarray} 
\label{Q10}
U = V^{-1}DV .
\end{eqnarray}
With the explicit representations
\begin{eqnarray} 
\label{Q11}
D = e^{i\xi} e^{i\rho\sigma_3}
=\left( {\matrix{{e^{i\theta _+}}&0\cr
0&{e^{i\theta _-}}\cr
}} \right) ,
\ \ \
\theta_\pm := \xi\pm\rho ,
\ \ \ {\rm and} \ \ \
V = 
e^{i{\mu \over 2}\sigma_2} e^{i{\nu \over 2}\sigma_3},
\end{eqnarray}
one can show easily that
with
$\sigma_V$ 
$:=$
$e^{-i{\nu \over 2}\sigma_3}$
$e^{-i{\mu \over 2}\sigma_2}$
$e^{i{\nu \over 2}\sigma_3} \sigma_3$
$= \sigma_V^{-1}$,
one has
\begin{eqnarray} 
\label{Q13}
U=\sigma_V D \sigma_V 
\end{eqnarray}
which is of the type (\ref{P9}).
One can therefore conclude that 
[A] the spectrum of the system described by $H_U$ is uniquely 
determined by the {\it eigenvalue} of $U$, and
[B] a point interaction
characterized by $U$ possesses the
{\rm isospectral subfamily}  
\begin{eqnarray} 
\Omega_{iso} := \left\{  H_{ V^{-1} D V } \vert 
V \in SU(2)  \right\} ,
\end{eqnarray}
which
is homeomorphic
to the 2-sphere specified by the polar angles $(\mu , \nu)$.
\begin{eqnarray} 
\Omega_{iso}
= \left\{ (\mu, \nu) \vert
\mu\in [0,\pi], \nu \in [0,2\pi) \right\}
\simeq S^2 .
\end{eqnarray}
There is of course an obvious exception to this
for the case 
of $D$ $\propto I$, in which case, $\Omega_{iso}$ 
consists only of $D$ itself.

To see the structure of the spectral space, {\it i.e.} the 
part of parameter space $U(2)$ that determines the distinct
spectrum of the system, 
it is convenient to make the spectrum 
of the system discrete.  Here, for simplicity, 
we consider
the line $x \in [-l,l]$ with Dirichlet 
boundary, $\varphi(-l)$ $= \varphi(l)$ $ = 0$.
%
%
One then has
\begin{eqnarray} 
\label{P22} 
V 
  \left( {\matrix{{\varphi (0_+)}\cr
                  {\varphi (0_-)}\cr}
         } 
  \right) 
= \sin{kl} \Phi_0,
\qquad
V
  \left( {\matrix{{ \varphi' (0_+)}\cr
                  {-\varphi' (0_-)}\cr}
         } 
  \right) 
= k\cos{kl} \Phi_0,
\end{eqnarray}
with some common constant vector $\Phi_0$.
From (4), we obtain
\begin{eqnarray} 
\label{R15}
1 + k L_0 \cot{kl} \cot{\theta_+\over 2} = 0,
\qquad
1 + k L_0 \cot{kl} \cot{\theta_-\over 2} = 0.
\end{eqnarray}
This means that the spectrum of the system is effectively 
split into that of two separate systems of same structure, 
each characterized by the parameters $\theta_+$ 
and $\theta_-$.   So the spectra of the system is
uniquely determined by two angular parameters 
$\{\theta_+, \theta_-\}$.

The entire parameter space 
$\Omega$ = $\{\theta_+, \theta_-, \mu, \nu \}$ is a product of 
spectral space  
$\Omega_{sp}$ = $\{\theta_+, \theta_- \}$ 
which is homeomorphic to
the 2-torus $T^2 = S^1 \times S^1$ 
and the isospectral space
$\Omega_{iso}$ = $\{ \mu, \nu \}$ $\simeq S^2$ (See Fig. 1a).
%
Note, however, that this parameter space provides a double 
covering for the family of point inteactions $\Omega \simeq U(2)$ 
due to the arbitrariness in the 
interchange $\theta_+ \leftrightarrow \theta_-$.  
Accordingly, two systems with interchanged
values for $\theta_+$ and $\theta_-$ are isospectral.
So the space of distinct spectra $\Sigma$ 
is the torus 
$T^2 = \{ (\theta_+,\theta-) | \theta_\pm \in [0,2\pi)\}$
subject to the identification    
$(\theta_+,\theta_-)$ $\equiv (\theta_-,\theta_+)$.
Thus we have
\begin{eqnarray} 
\label{R30} 
\Sigma :=\{ Spec(H_U) | U \in \Omega \} = T^2/{\bf Z}_2 , 
\end{eqnarray}
which is homeomorphic to a M{\"o}bius strip with boundary
(Fig. 1b).

To relate the non-trivial topological structure found here and the
exotic quantum phenomena we have alluded to in the introduction, the
readers are referred to other publications.\cite{TF00,CF01,TF01}
Here we simply observe that the homotopy 
$\pi_1(T^2)={\bf Z} \times {\bf Z}$ is behind the double spiral 
anholonomy,\cite{CH98} and the isospectral family $S^2$ is 
the generalization
of the duality\cite{CS99} found earlier.
\\

This work has been supported in part by 
the Monbu-Kagakusho Grant-in-Aid for Scientific Research 
(Nos. (C)11640301 and (C)13640413).

\end{document}